\documentclass[Physsubmission, Phys]{SciPost}

\hypersetup{
    colorlinks,
    linkcolor={red!50!black},
    citecolor={blue!50!black},
    urlcolor={blue!80!black}
}

\usepackage[bitstream-charter]{mathdesign}
\DeclareSymbolFont{usualmathcal}{OMS}{cmsy}{m}{n}
\DeclareSymbolFontAlphabet{\mathcal}{usualmathcal}

\def\beq{\begin{equation}}
\def\eeq{\end{equation}}
\def\beqa{\begin{eqnarray}}
\def\eeqa{\end{eqnarray}}

\begin{document}

\begin{center}{\Large \textbf{
Three-loop soft anomalous dimensions for top-quark processes}}
\end{center}

\begin{center}
Nikolaos Kidonakis\textsuperscript{$\star$}
\end{center}

\begin{center}
Kennesaw State University, Kennesaw, GA 30144, USA
\\
* nkidonak@kennesaw.edu
\end{center}

\begin{center}
\today
\end{center}

\definecolor{palegray}{gray}{0.95}
\begin{center}
\colorbox{palegray}{
  \begin{tabular}{rr}
  \begin{minipage}{0.1\textwidth}
    \includegraphics[width=22mm]{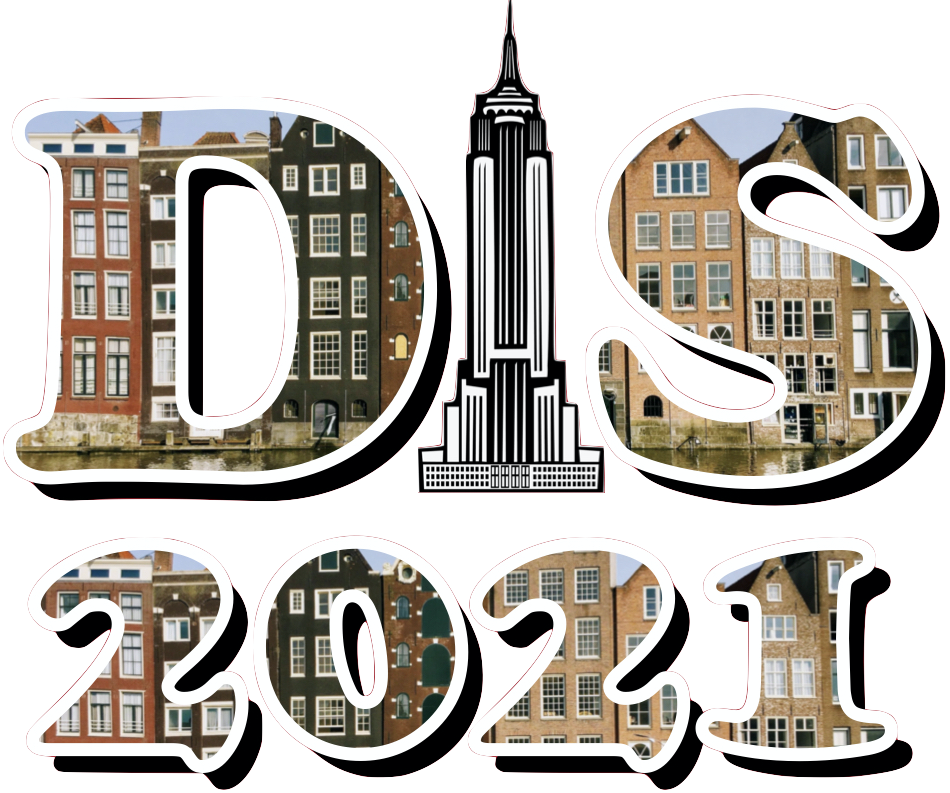}
  \end{minipage}
  &
  \begin{minipage}{0.75\textwidth}
    \begin{center}
    {\it Proceedings for the XXVIII International Workshop\\ on Deep-Inelastic Scattering and
Related Subjects,}\\
    {\it Stony Brook University, New York, USA, 12-16 April 2021} \\
    \doi{10.21468/SciPostPhysProc.?}\\
    \end{center}
  \end{minipage}
\end{tabular}
}
\end{center}

\section*{Abstract}
{\bf
I present results for soft anomalous dimensions through three loops for several processes involving the production of top quarks. In particular, I discuss single-top and top-pair production. I also present some numerical results for double-differential distributions in $t{\bar t}$ production through approximate N$^3$LO.
}

\section{Introduction}

The inclusion of soft-gluon corrections in theoretical predictions for top-quark processes is required for better accuracy,  and it involves calculations of soft anomalous dimensions. The first calculations at one loop were done in the mid 90's \cite{NKGS}, but two-loop calculations appeared much later. The current state-of-the-art has been extended to three loops for some processes.

For partonic processes $f_{1}(p_1)\, + \, f_{2}\, (p_2) \rightarrow t(p_t)\, + \, X $,
we define a kinematical threshold variable $s_4=s+t+u-\sum_i m_i^2$ where $s=(p_1+p_2)^2$, $t=(p_1-p_t)^2$, and $u=(p_2-p_t)^2$. At partonic threshold $s_4 \rightarrow 0$, and the soft-gluon corrections involve logarithms of the form $[\ln^k(s_4/m_t^2)/s_4]_+$ with $k \le 2n-1$ at perturbative order $\alpha_s^n$. 

We define transforms of the partonic cross section as 
${\hat \sigma}(N)=\int (ds_4/s) \; e^{-N s_4/s} {\hat \sigma}(s_4)$, 
with transform variable $N$. The 
factorized expression for the cross section is \cite{NKGS}
\beq
\sigma^{f_1 f_2\rightarrow tX}(N)
= H^{f_1 f_2\rightarrow tX} \, 
S^{f_1 f_2 \rightarrow tX}\left(\frac{m_t}{N \mu_F}\right)
\, \psi_1 \left(N_1,\mu_F\right) \, 
\psi_2 \left(N_2,\mu_F \right) \, 
\prod_i J_i \left(N,\mu_F \right)
\eeq
where
$H^{f_1 f_2\rightarrow tX}$ is an $N$-independent hard function, $S^{f_1 f_2\rightarrow tX}$ is a soft function \cite{NKGS}, while the $\psi_i$ and $J_i$ describe collinear emission from initial- and final-state particles \cite{GS}.

The soft function $S^{f_1 f_2\rightarrow tX}$ satisfies the renormalization group equation
\beq
\left(\mu_R \frac{\partial}{\partial \mu_R}
+\beta(g_s)\frac{\partial}{\partial g_s}\right)\,S^{f_1 f_2\rightarrow tX}
=-\Gamma_S^{\dagger \, f_1 f_2\rightarrow tX} \, S^{f_1 f_2\rightarrow tX}
-S^{f_1 f_2\rightarrow tX} \, \Gamma_S^{f_1 f_2\rightarrow tX}
\eeq
where the soft anomalous dimension $\Gamma_S^{f_1 f_2\rightarrow tX}$ controls the evolution of $S^{f_1 f_2\rightarrow tX}$, which gives the exponentiation of logarithms of $N$ in the resummed cross section. The resummation of these soft corrections at NNLL accuracy requires knowledge of two-loop soft anomalous dimensions while at N$^3$LL accuracy it requires three-loop soft anomalous dimensions. The resummed cross sections may be expanded at finite order and produce, upon inversion to momentum space, physical predictions.

A recent review of calculations of cusp and soft anomalous dimensions for many processes can be found in Ref. \cite{NKuni} (see also \cite{NK2loop,GHKM,NK3loopcusp} for the cusp, \cite{NKsingletop,NKsch,NKtWH,NKtch,NK3loop} for single top, and \cite{NKGS,NKttbar} for $t{\bar t}$).

\section{Cusp anomalous dimension}

The cusp anomalous dimension, $\Gamma_{\rm cusp}$, involves two eikonal lines, and it is a basic ingredient in calculations of soft anomalous dimensions for partonic processes. For two lines with momenta $p_i$ and $p_j$, the cusp angle is $\theta=\cosh^{-1}(p_i\cdot p_j/\sqrt{p_i^2 p_j^2})$, and we write the perturbative series $\Gamma_{\rm cusp}=\sum_{n=1}^{\infty} \left(\frac{\alpha_s}{\pi}\right)^n \Gamma^{(n)}_{\rm cusp}$. At one loop, $\Gamma_{\rm cusp}^{(1)}=C_F (\theta \coth\theta-1)$. In the case of two heavy quarks, this can be written in terms of the speed $\beta=\tanh(\theta/2)$ as
\beq
\Gamma_{\rm cusp}^{(1) \, \beta}=-C_F\left[\frac{(1+\beta^2)}{2\beta}
\ln\frac{(1-\beta)}{(1+\beta)}+1\right] \, .
\eeq

At two loops, we have \cite{NK2loop}
\beqa
\Gamma_{\rm cusp}^{(2)}&=& K_2 \, \Gamma_{\rm cusp}^{(1)}
+\frac{1}{2}C_F C_A \left\{1+\zeta_2+\theta^2 
-\coth\theta\left[\zeta_2\theta+\theta^2
+\frac{\theta^3}{3}+{\rm Li}_2\left(1-e^{-2\theta}\right)\right] \right. 
\nonumber \\ && \hspace{20mm} \left.
{}+\coth^2\theta\left[-\zeta_3+\zeta_2\theta+\frac{\theta^3}{3}
+\theta \, {\rm Li}_2\left(e^{-2\theta}\right)
+{\rm Li}_3\left(e^{-2\theta}\right)\right] \right\} 
\eeqa
where $K_2=C_A \left(\frac{67}{36}-\frac{\zeta_2}{2}\right)-\frac{5}{18} n_f$.
This can be written in terms of $\beta$ and denoted as $\Gamma_{\rm cusp}^{(2) \, \beta}$.

The three-loop result \cite{GHKM,NK3loopcusp} can be written as \cite{NK3loopcusp,NKuni} 
\beq
\Gamma_{\rm cusp}^{(3)}= K_3 \Gamma_{\rm cusp}^{(1)}
+2 \, K_2 \left(\Gamma_{\rm cusp}^{(2)}-K_2 \, \Gamma_{\rm cusp}^{(1)}\right)
+C^{(3)}
\eeq
where $K^{(3)}$ and $C^{(3)}$ have long expressions. Again, the result can be expressed in terms of $\beta$.

If eikonal line $i$ represents a massive quark, with mass $m_i$, and eikonal 
line $j$ a massless quark, then we find simpler expressions.
At one loop, $\Gamma_{\rm cusp}^{(1) \, m_i}=C_F [\ln(2 p_i \cdot p_j/(m_i \sqrt{s}))-1/2]$. 
At two loops \cite{NKtWH,NK3loop},  $\Gamma^{(2) \, m_i}_{\rm cusp}=K_2 \, \Gamma_{\rm cusp}^{(1) \, m_i}+C_F C_A (1-\zeta_3)/4$.
At three loops \cite{NK3loop},
\beq
\Gamma^{(3) \, m_i}_{\rm cusp}=K_3 \, \Gamma_{\rm cusp}^{(1) \, m_i}
+ \frac{1}{2} K_2 C_F C_A (1-\zeta_3)
+C_F C_A^2\left(-\frac{1}{4}+\frac{3}{8}\zeta_2-\frac{\zeta_3}{8}-\frac{3}{8}\zeta_2 \zeta_3+\frac{9}{16} \zeta_5\right) \, .
\eeq

If both eikonal lines are massless, then
$\Gamma_{\rm cusp}^{\rm massless}=C_F \ln(2 p_i \cdot p_j/s) 
\sum_{n=1}^{\infty} (\alpha_s/\pi)^n K_n$.

\section{Single-top production}

Next, we discuss various single-top production processes \cite{NKsingletop,NKsch,NKtWH,NKtch,NK3loop}.

\subsection{Single-top $t$-channel production}

The soft anomalous dimension for $t$-channel single-top production, ${\Gamma}_S^{bq \to tq'}$, is a $2 \times 2$ matrix in color space. We use a $t$-channel singlet-octet color basis.
The one-loop \cite{NKsingletop,NKtch,NK3loop} and two-loop \cite{NKtch,NK3loop} results are well known.

At three loops, we have
\beqa
&&\Gamma_{S \, 11}^{(3) bq \to tq'}= K_3 \, \Gamma_{S \, 11}^{(1) bq \to tq'}
+ \frac{1}{2} K_2 C_F C_A (1-\zeta_3) 
+C_F C_A^2\left(-\frac{1}{4}+\frac{3}{8}\zeta_2-\frac{\zeta_3}{8}-\frac{3}{8}\zeta_2 \zeta_3+\frac{9}{16} \zeta_5\right)
\nonumber \\ &&
\Gamma_{S \, 12}^{(3) bq \to tq'}= K_3 \, \Gamma_{S \, 12}^{(1) bq \to tq'} + X_{S \, 12}^{(3) bq \to tq'} \, , \quad \quad   
\Gamma_{S \, 21}^{(3) bq \to tq'}= K_3 \, \Gamma_{S \, 21}^{(1) bq \to tq'} + X_{S \, 21}^{(3) bq \to tq'}
\nonumber \\ &&
\Gamma_{S \, 22}^{(3) bq \to tq'}= K_3 \, \Gamma_{S \, 22}^{(1) bq \to tq'}
+ \frac{1}{2} K_2 C_F C_A (1-\zeta_3) 
\nonumber \\ && \hspace{20mm}
{}+C_F C_A^2\left(-\frac{1}{4}+\frac{3}{8}\zeta_2-\frac{\zeta_3}{8}-\frac{3}{8}\zeta_2 \zeta_3+\frac{9}{16} \zeta_5\right) + X_{S \, 22}^{(3) bq \to tq'} \, .
\eeqa

The first element, i.e. the "11" element, of the matrix at three loops was calculated in \cite{NK3loop}. Due to the relatively simple color structure of the hard matrix for this process, it is the only three-loop element that contributes to the N$^3$LO soft-gluon corrections. Here we have also provided three-loop results for the other three matrix elements up to unknown terms from four-parton correlations, which are denoted as $X^{(3) bq \to tq'}_S$ in the above equation. 

\subsection{Single-top $s$-channel production}

We continue with results for the $s$-channel, for which ${\Gamma}_S^{q{\bar q}' \to t{\bar b}}$ is a $2 \times 2$ matrix, and we use an $s$-channel singlet-octet color basis. The one-loop \cite{NKsingletop,NKsch,NK3loop} and two-loop \cite{NKsch,NK3loop} results are, again, well known.

At three loops, we have
\beqa
&& \Gamma_{S \, 11}^{(3) q{\bar q}' \to t{\bar b}}= K_3 \, \Gamma_{S \, 11}^{(1) q{\bar q}' \to t{\bar b}}
+\frac{1}{2} K_2 C_F C_A (1-\zeta_3)  
+C_F C_A^2\left(-\frac{1}{4}+\frac{3}{8}\zeta_2-\frac{\zeta_3}{8}-\frac{3}{8}\zeta_2 \zeta_3+\frac{9}{16} \zeta_5\right)
\nonumber \\ &&
\Gamma_{S \, 12}^{(3) q{\bar q}' \to t{\bar b}}=K_3 \, \Gamma_{S \, 12}^{(1) q{\bar q}' \to t{\bar b}} + X_{S \, 12}^{(3) q{\bar q}' \to t{\bar b}} \, , 
\quad \quad
\Gamma_{S \, 21}^{(3) q{\bar q}' \to t{\bar b}}= K_3 \, \Gamma_{S \, 21}^{(1) q{\bar q}' \to t{\bar b}} + X_{S \, 21}^{(3) q{\bar q}' \to t{\bar b}} 
\nonumber \\ &&
\Gamma_{S \, 22}^{(3) q{\bar q}' \to t{\bar b}}= K_3 \, \Gamma_{S \, 22}^{(1) q{\bar q}' \to t{\bar b}}
+\frac{1}{2} K_2 C_F C_A (1-\zeta_3)
\nonumber \\ && \hspace{20mm}
{}+C_F C_A^2\left(-\frac{1}{4}+\frac{3}{8}\zeta_2-\frac{\zeta_3}{8}-\frac{3}{8}\zeta_2 \zeta_3+\frac{9}{16} \zeta_5\right) + X_{S \, 22}^{(3) q{\bar q}' \to t{\bar b}}
\eeqa

Again, the "11" element of the matrix at three loops was calculated in \cite{NK3loop} and is the only three-loop element to contribute to the N$^3$LO soft-gluon corrections. We have also provided in the above equation three-loop results for the other three matrix elements up to unknown terms from four-parton correlations, which are denoted as $X^{(3) q{\bar q}' \to t{\bar b}}_S$. 

\subsection{Associated $tW$ production}

The soft anomalous dimension for $tW$ production has only one element (not a matrix). It is known at one loop \cite{NKsingletop,NKtWH}, two-loops \cite{NKtWH}, and three loops \cite{NK3loop}. The three-loop result is \cite{NK3loop}
\beq
\Gamma_S^{(3) bg \to tW}=K_3 \, \Gamma_S^{(1) bg \to tW}+\frac{1}{2} K_2 C_F C_A (1-\zeta_3)
+C_F C_A^2\left(-\frac{1}{4}+\frac{3}{8}\zeta_2-\frac{\zeta_3}{8}-\frac{3}{8}\zeta_2 \zeta_3+\frac{9}{16} \zeta_5\right) \, .
\eeq

\section{Top-antitop pair production}

We continue with top-antitop pair production which can proceed via the $q{\bar q} \to t{\bar t}$ and the $gg \to t{\bar t}$ channels.

In the $q{\bar q} \to t{\bar t}$ channel, 
${\Gamma}_S^{q{\bar q} \to t{\bar t}}$ is a $2 \times 2$ matrix, and we use an $s$-channel singlet-octet color basis. Here we will concentrate on the "22" matrix element which at one-loop contributes already to the soft-gluon corrections at NLO.
At one loop, this element is \cite{NKGS,NKttbar}
\beq
\Gamma^{(1) q{\bar q} \to t{\bar t}}_{22}=\left(1-\frac{C_A}{2C_F}\right)
\Gamma_{\rm cusp}^{(1) \, \beta} 
+4C_F \ln\left(\frac{t-m_t^2}{u-m_t^2}\right)
-\frac{C_A}{2}\left[1+\ln\left(\frac{s m_t^2 (t-m_t^2)^2}{(u-m_t^2)^4}\right)\right] \eeq
while at two loops it is \cite{NKttbar,NKuni}
\beq
\Gamma^{(2) q{\bar q} \to t{\bar t}}_{22}=
K_2 \Gamma^{(1) q{\bar q} \to t{\bar t}}_{22}
+\left(1-\frac{C_A}{2C_F}\right)
\left(\Gamma_{\rm cusp}^{(2) \, \beta}-K_2 \Gamma_{\rm cusp}^{(1) \, \beta}\right)
+\frac{C_A^2}{4}(1-\zeta_3) \, .
\eeq
At three loops, we find the following expression: 
\beqa
\Gamma_{S \, 22}^{(3) q{\bar q} \to t{\bar t}}&=&
K_3 \, \Gamma_{S \, 22}^{(1) q{\bar q} \to t{\bar t}}
+\left(1-\frac{C_A}{2C_F}\right) \left(\Gamma_{\rm cusp}^{(3) \, \beta}-K_3 \Gamma_{\rm cusp}^{(1) \, \beta}\right)+\frac{K_2}{2}C_A^2(1-\zeta_3)
\nonumber \\ &&
{}+C_A^3\left(-\frac{1}{4}+\frac{3}{8}\zeta_2-\frac{\zeta_3}{8}-\frac{3}{8}\zeta_2\zeta_3+\frac{9}{16}\zeta_5\right)+X_{S \, 22}^{(3) q{\bar q} \to t{\bar t}}
\eeqa
where $X_{S \, 22}^{(3) q{\bar q} \to t{\bar t}}$ denotes unknown three-loop contributions from four-parton correlations.
The other three-loop matrix elements are not fully known either but have analogous structure to that at two loops (see also \cite{NKuni}).

In the $gg \to t{\bar t}$ channel, $\Gamma_{S \, 22}^{gg \to t{\bar t}}$ is a $3\times3$ matrix, and we use the color basis $c_1=\delta^{ab} \delta_{12}$, $c_2= d^{abc} T^c_{12}$, $c_3= if^{abc} T^c_{12}$. At one loop for $gg \to t{\bar t}$, the "22" matrix element is \cite{NKGS,NKttbar}
\beq
\Gamma_{S \, 22}^{(1) gg \to t{\bar t}}=\left(1-\frac{C_A}{2C_F}\right) \Gamma_{\rm cusp}^{(1) \, \beta} 
+\frac{C_A}{2}\left[\ln\left(\frac{(t-m_t^2)(u-m_t^2)}{s\, m_t^2}\right)-1\right]
\eeq
while at two loops it is \cite{NKttbar,NKuni}
\beq
\Gamma_{S \, 22}^{(2) gg \to t{\bar t}}= K_2 \, \Gamma_{S \, 22}^{(1) gg \to t{\bar t}}
+\left(1-\frac{C_A}{2C_F}\right) \left(\Gamma_{\rm cusp}^{(2) \, \beta}-K_2 \Gamma_{\rm cusp}^{(1) \, \beta}\right)+\frac{C_A^2}{4}(1-\zeta_3) \, .
\eeq
At three loops, we find the expression 
\beqa
\Gamma_{S \, 22}^{(3) gg \to t{\bar t}}&=&
K_3 \, \Gamma_{S \, 22}^{(1) gg \to t{\bar t}}
+\left(1-\frac{C_A}{2C_F}\right) \left(\Gamma_{\rm cusp}^{(3) \, \beta}-K_3 \Gamma_{\rm cusp}^{(1) \, \beta}\right)+\frac{K_2}{2}C_A^2(1-\zeta_3)
\nonumber \\ &&
{}+C_A^3\left(-\frac{1}{4}+\frac{3}{8}\zeta_2-\frac{\zeta_3}{8}-\frac{3}{8}\zeta_2\zeta_3+\frac{9}{16}\zeta_5\right)+X_{S \, 22}^{(3) gg \to t{\bar t}}
\eeqa
where $X_{S \, 22}^{(3) gg \to t{\bar t}}$ denotes unknown three-loop contributions from four-parton correlations.

\begin{figure}[ht]
\centering
\includegraphics[width=0.49\textwidth]{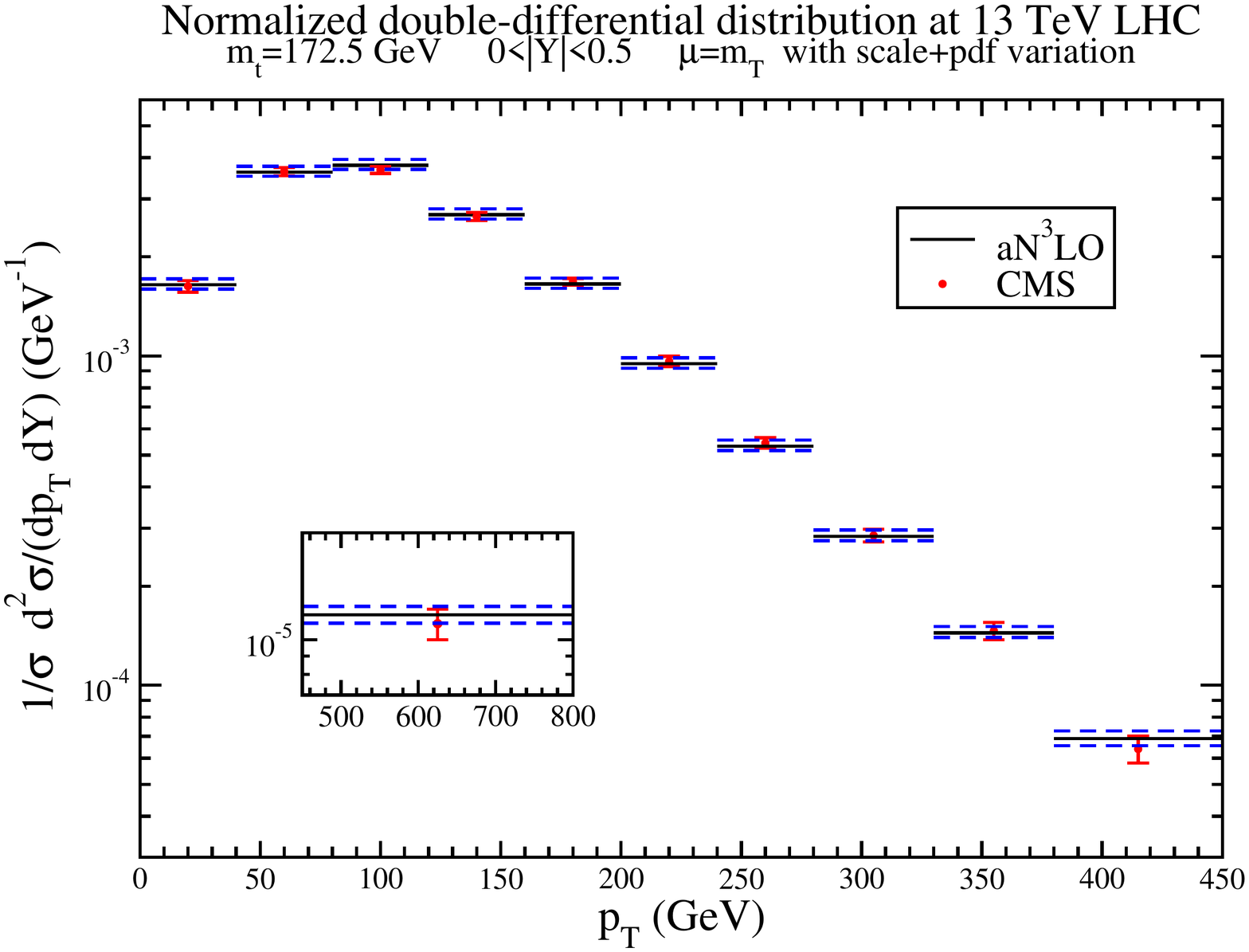}
\includegraphics[width=0.49\textwidth]{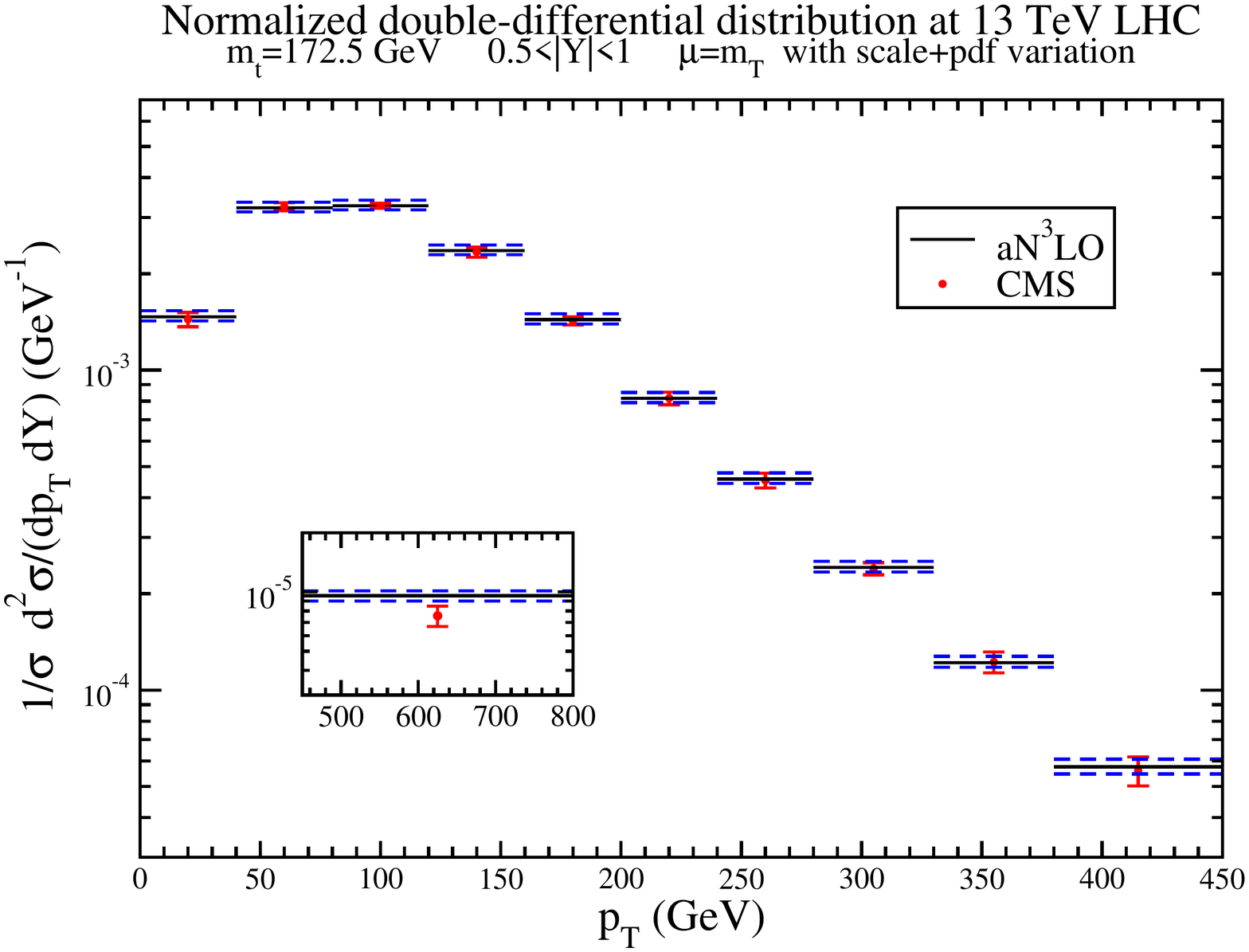}
\caption{Top-quark double-differential distributions.}
\label{doublediff}
\end{figure}

As an application of the soft-gluon formalism at NNLL accuracy, in Figure 1 we show top-quark double-differential distributions in $p_T$ and rapidity with soft-gluon corrections through approximate N$^3$LO \cite{NKdoublediff}. The theoretical predictions describe very well the CMS data at 13 TeV \cite{CMS}.

\section{Conclusion}

I have presented results for soft anomalous dimensions at one, two, and three loops.
The cusp anomalous dimension was discussed first followed by results for the soft
anomalous dimensions in single-top production and in top-antitop pair production.

\paragraph{Funding information}
This material is based upon work supported by the National Science Foundation under Grant No. PHY 1820795.

\end{document}